\documentclass[11pt,a4paper]{article}

\usepackage{jheppub_mod}

\usepackage{axodraw}
\usepackage{dcolumn,booktabs}
\usepackage{bm,amsfonts,mathrsfs,ulem}

\newcommand{\M}{\mathcal{M}}
\renewcommand{\B}{\mathcal{B}}
\renewcommand{\Im}{\text{Im}}
\renewcommand{\Re}{\text{Re}}
\renewcommand{\vec}[1]{\mathbf{#1}}
\newcommand{\nnnl}{\nonumber\\}
\newcommand{\lam}{\lambda}
\newcommand{\kap}{\kappa}

\title{Improving the Hadron Physics  of Non-Standard-Model Decays: Example
Bounds on R-parity Violation}

\author[a,b]{J.~T.~Daub,}
\author[a,c]{H.~K.~Dreiner,}
\author[d]{C.~Hanhart,}
\author[a,b]{B.~Kubis,}
\author[a,b,d]{and U.-G.~Mei{\ss}ner}

\affiliation[a]{
Bethe Center for Theoretical Physics, Universit\"at Bonn, Nu\ss allee 12,
D-53115 Bonn, Germany}

\affiliation[b]{
Helmholtz-Institut f\"ur Strahlen- und Kernphysik (Theorie),
Universit\"at  Bonn,
Nu{\ss}allee 14--16, D-53115 Bonn, Germany}

\affiliation[c]{
Physikalisches Institut, Universit\"at Bonn, Nu\ss allee 12,
D-53115 Bonn, Germany}

\affiliation[d]{
Institut f\"ur Kernphysik, Institute for Advanced Simulation 
and J\"ulich Center for Hadron Physics,
Forschungszentrum J\"ulich, D-52425 J\"{u}lich, Germany}

\emailAdd{daub@hiskp.uni-bonn.de}
\emailAdd{dreiner@uni-bonn.de}
\emailAdd{c.hanhart@fz-juelich.de}
\emailAdd{kubis@hiskp.uni-bonn.de}
\emailAdd{meissner@hiskp.uni-bonn.de}

\abstract{
  Using the example of selected decays driven by R-parity-violating
  supersymmetric operators, we demonstrate how strong final-state
  interactions can be controlled quantitatively with high precision, 
  thus allowing for a more accurate extraction of effective parameters
  from data. In our examples we focus on the lepton-flavor-violating
  decays $\tau\to\mu\pi^+\pi^-$. In R-parity violation these can arise
  due to the product of two couplings. We find bounds that are an order of
  magnitude stronger than previous ones.
}


\begin{document}

\maketitle

\section{Introduction}
If weak-scale supersymmetry~\cite{Nilles:1983ge} is the solution to
the hierarchy problem there must be new supersymmetric particles with
masses below $ \mathcal{O}(5\,\mathrm{TeV})$, which are accessible to
the LHC. There are basically two forms of the minimal supersymmetric
Standard Model, with light supersymmetric fields, distinguished by
their superpotential.  The most widely studied case is R-parity
conservation, where the symmetries of the supersymmetric Standard
Model are extended to include the discrete multiplicative symmetry
R-parity.  This renders the proton stable in the theory and the
resulting renormalizable superpotential is given by
\begin{equation}
W_{\mathrm{MSSM}}=\epsilon_{ab}
\left[(h_e)_{ij} L_i^a H_d^b \bar E_j + (h_d)_{ij} Q_i^a 
H_d^b\bar D_j + (h_u)_{ij} Q_i^a H_u^b\bar U_j+\mu H_d^aH_u^b\right]\,.
\label{superpot}
\end{equation}
Here $L,\,Q, H_d,\,H_u$ are the lepton, quark, and Higgs SU(2)$_L$
doublet left chiral superfields, respectively. $\bar E,\,\bar D,\,\bar
U$ are the corresponding SU(2)$_L$ singlet lepton and quark left
chiral superfields. $i,j,k\in\{1,2,3\}$ are generation indices and $a,
\,b\in\{1,2\}$ are SU(2)$_L$ gauge indices. The $h_{f=e,d,u}$ are
dimensionless Yukawa couplings and $\mu$ is the Higgs mass mixing
term. Note that if non-renormalizable terms are allowed then for
example the R-parity conserving superpotential term $QQQL$ leads to a
dimension-five proton decay operator~\cite{Nilles:1983ge}. This is
thus only suppressed by one power of the large mass scale. In this
case proton hexality~\cite{Dreiner:2005rd,Dreiner:2003yr} is the
appropriate symmetry.  It leads to the same low-energy superpotential
given in Eq.~\eqref{superpot}, but prohibits all dimension-five proton
decay operators.

Alternatively, the proton is also stable for the discrete $\mathbb{Z}_3$ 
symmetry baryon triality~\cite{Ibanez:1991pr}. The resulting
renormalizable superpotential is~\cite{Dreiner:1997uz} 
\begin{equation}
W_{\mathrm{B}_3}=W_{\mathrm{MSSM}}+\epsilon_{ab}\left[\lam_{ijk} 
L_i^a L_j^b\bar E_k + \lam'_{ijk} L_i^aQ_j^b 
\bar D_k +\kap_i L_i^a H_u^b\right]\,.
\label{rpv-superpot}
\end{equation}
Here the $\lam_{ijk},\,\lam'_{ijk}$ are dimensionless couplings and
the $\kap_i$ have dimension mass. Both baryon triality and proton
hexality are discrete gauge anomaly-free in the sense
of Refs.~\cite{Dreiner:2005rd,Ibanez:1991pr,Krauss:1988zc}.  At any given
energy scale the $\kap_i$ can be rotated to zero, by a transformation
in $(L_i,\,H_d)$ space~\cite{Hall:1983id,Dreiner:2003hw}. Since we
perform our computations at a fixed low-energy scale, we shall focus
on the case $\kappa_i=0$ in the following.

An important feature of the additional terms in
Eq.~\eqref{rpv-superpot} is that they violate lepton number and flavor.
Correspondingly there is a large set of bounds on the couplings 
$\lambda_{ijk}$ and $\lambda^\prime_{ijk}$. This was first considered
in Ref.~\cite{Hall:1983id}, including also possible contributions to
rare meson decays. A more systematic approach was taken in
Ref.~\cite{Barger:1989rk}, considering bounds on all couplings
individually.  Since then many bounds have been set from the decays of
mesons, including also bounds on products of 
operators~\cite{Bhattacharyya:1995pq,Agashe:1995qm,Bhattacharyya:1995cg,Altarelli:1997ce,Dreiner:1997cd,Kim:1997rr,Allanach:1999ic,herbi02,herbi07}. In Ref.~\cite{Barger:1989rk}, for
example, the authors considered the R-parity-violating contributions
to
\begin{equation}
R_\pi\equiv \frac{\Gamma(\pi\to e\nu)}{\Gamma(\pi\to \mu\nu)}\,.
\end{equation}
An operator $\lam'_{21k} L_2Q_1\bar D_k$ contributes only to
${\Gamma(\pi\to \mu\nu)}$, thus modifying $R_\pi$. In computing the
exact contribution to ${\Gamma(\pi\to \mu\nu)}$, one uses the definition
\begin{equation}
\langle0|\bar u \gamma^\mu P_L d |\pi^-(q)\rangle\equiv -\frac{i}{\sqrt{2}}\,f_\pi q^\mu 
\end{equation}
of the pion decay constant $f_\pi$ for the $V-A$ current, 
and the current-algebra approximation
\begin{equation}
\langle0|\bar u P_L d |\pi^-(q)\rangle= \sqrt{2} i\,\frac{M_\pi^2} {m_u+m_d}f_\pi 
\end{equation}
for the chiral (pseudo)scalar coupling. Here, $M_\pi$ denotes the
charged pion mass, $m_{u,d}$ are the first-generation quark masses,
and we make use of the left- and right-handed projection operators $P_{L/R}=(1\mp\gamma_5)/2$.
Using the experimental value including the error of $R_\pi$~\cite{PDG}, 
which agrees with the Standard Model, results in a bound
on $\lambda'_{21k}$. This assumes that $L_2Q_1\bar D_k$ is the sole
new operator contributing.

More recently Herrero and collaborators have studied $\tau$ decays
in various R-parity conserving supersymmetric
models~\cite{Arganda:2008jj,herrero09}, for example the decays
\begin{equation}
\tau^\pm\rightarrow \mu^\pm + 
f_0  \,/\,
\pi^+\pi^-,\,\pi^0\pi^0 \,/\,
K^+K^-,\,K^0\bar K^0 \,/\,
\eta,\,\eta' \,/\,
\rho^0,\,\phi ~.
\end{equation}
The authors go beyond the simple current algebra approximation and
employ chiral perturbation theory and resonance chiral theory~\cite{resChPT}. 
They thus dramatically improve the precision of
the computation and therefore also the resulting bounds on new
physics. The purpose of this paper is to refine these techniques
further---in particular, we include both the scalar form factors and
the vector form factor, model-independently. In addition, we will also
discuss the case of R-parity violation. Specifically we shall focus on
the decay
\begin{equation}
\tau^\pm\rightarrow\mu^\pm \pi^+\pi^-
\end{equation}
to clarify our method. In terms of R-parity-violating operators, this
decay receives contributions via the parton level processes
\begin{equation}
\tau^\pm\rightarrow\mu^\pm u_i\bar u_k  \qquad \mathrm{and}\qquad\tau^\pm\rightarrow\mu^\pm d_j\bar d_k ~.
\end{equation}
Combining the operators $\epsilon_{ab}\lam_{ijk} L_i^a L_j^b\bar E_k$
and $\epsilon_{ab} \lam'_{ijk} L_i^aQ_j^b \bar D_k$ from the
superpotential in Eq.~\eqref{rpv-superpot} and integrating out the heavy
intermediate scalar fermion, we obtain several independent
contributions to the decay $\tau^-\to\mu^-+2$~quarks
(cf.\ Appendix~\ref{app:effops}):
\begin{enumerate}
\item[(a)] Combining $\epsilon_{ab} \lam'_{3ij} L_3^aQ_i^b  \bar D_j$
and $[\epsilon_{ab} \lam'_{2kj} L_2^aQ_k^b  \bar D_j]^\dagger$ corresponds
to $\tau^-\to\mu^-\bar u_i u_k$ with
\begin{equation}
\mathscr{L}_{\mathrm{eff}}=\frac{1}{2}
\frac{\lambda'_{3ij}\lambda^{\prime *}_{2kj}}{m^2_{\tilde d_j}}
(\bar u_k\gamma^\alpha P_L u_i)(\bar\mu \gamma_\alpha P_L \tau) ~.\label{eq:a}
\end{equation}
\item[(b)] Combining
$\epsilon_{ab} \lam'_{3ij} L_3^aQ_i^b  \bar D_j$
and $\left[\epsilon_{ab} \lam'_{2i\ell} L_2^aQ_i^b  \bar D_\ell\right]^\dagger$
corresponds to $\tau^-\to\mu^- d_j \bar d_\ell$ with
\begin{equation}
\mathscr{L}_{\mathrm{eff}}=\frac{1}{2}
\frac{\lambda'_{3ij}\lambda^{\prime *}_{2i\ell}}{m^2_{\tilde u_i}}
(\bar d_j\gamma^\alpha P_R d_\ell)(\bar\mu \gamma_\alpha P_L \tau)  ~.\label{eq:b}
\end{equation}
\item[(c)] Combining $\epsilon_{ab} \lam_{3i2} L_3^aL_i^b \bar E_2$
  and $[\epsilon_{ab} \lam'_{ijk} L_i^aQ_j^b \bar D_k]^\dagger$
    corresponds to $\tau^-\to\mu^- d_j\bar d_k$ with
\begin{equation}
\mathscr{L}_{\mathrm{eff}}=\frac{1}{2}
\frac{\lambda_{3i2}\lambda^{\prime *}_{ijk}}{m^2_{\tilde \nu_i}}
(\bar d_j P_R d_k)(\bar\mu  P_L \tau) ~. \label{eq:c}
\end{equation}
\begin{figure}
\centering
    \scalebox{1.0}{
      \begin{picture}(190,140)(5,10)
	\ArrowLine(20,100)(70,100)                      
	\ArrowLine (100,120)(70,100)                     
	\ArrowLine (100,80)(130,100)                   
	\DashLine(100,80)(70,100){5}                      
	\ArrowLine(100,80)(130,60)                      
	\GCirc(70,100){2}{0}
	\GCirc(100,80){2}{0}	
	\put(30,110){\small$\boldsymbol{\tau^-}$}      
	\put(105,125){\small$\boldsymbol{\bar u_i}$}      
	\put(133,102){\small$\boldsymbol{\mu^-}$}         
	\put(77,80){\small$\boldsymbol{\tilde d_j}$}   
	\put(133,54){\small$\boldsymbol{u_k}$} 
	\put(80,50){\small$\boldsymbol{(a)}$} 
    \end{picture}}
    \scalebox{1.0}{
      \begin{picture}(190,140)(5,10)
	\ArrowLine(20,100)(70,100)                      
	\ArrowLine (100,120)(70,100)                     
	\ArrowLine (100,80)(130,100)                   
	\DashLine(100,80)(70,100){5}                      
	\ArrowLine(100,80)(130,60)                      
	\GCirc(70,100){2}{0}
	\GCirc(100,80){2}{0}	
	\put(30,110){\small$\boldsymbol{\tau^-}$}      
	\put(105,125){\small$\boldsymbol{d_j}$}      
	\put(133,102){\small$\boldsymbol{\mu^-}$}         
	\put(77,80){\small$\boldsymbol{\tilde u_i}$}   
	\put(133,54){\small$\boldsymbol{\bar d_\ell}$} 
	\put(80,50){\small$\boldsymbol{(b)}$} 
    \end{picture}}

\vspace{-1.5cm}

    \scalebox{1.0}{
      \begin{picture}(190,140)(5,10)
	\ArrowLine(20,100)(70,100)                      
	\ArrowLine (100,120)(70,100)                     
	\ArrowLine (100,80)(130,100)                   
	\DashLine(100,80)(70,100){5}                      
	\ArrowLine (100,80)(130,60)                    
	\GCirc(70,100){2}{0}
	\GCirc(100,80){2}{0}	
	\put(30,110){\small$\boldsymbol{\tau^-}$}      
	\put(105,125){\small$\boldsymbol{\mu^-}$}      
	\put(133,102){\small$\boldsymbol{d_j}$}         
	\put(77,80){\small$\boldsymbol{\tilde \nu_i}$}   
	\put(133,54){\small$\boldsymbol{\bar d_k}$} 
	\put(80,50){\small$\boldsymbol{(c)}$} 
    \end{picture}}
    \scalebox{1.0}{
      \begin{picture}(190,140)(5,10)
	\ArrowLine(70,100)(20,100)                      
	\ArrowLine (70,100)(100,120)                     
	\ArrowLine (130,100)(100,80)                  
	\DashLine(100,80)(70,100){5}                      
	\ArrowLine(130,60)(100,80)                    
	\GCirc(70,100){2}{0}
	\GCirc(100,80){2}{0}	
	\put(30,110){\small$\boldsymbol{\tau^-}$}      
	\put(105,125){\small$\boldsymbol{\mu^-}$}      
	\put(133,102){\small$\boldsymbol{d_j}$}         
	\put(77,80){\small$\boldsymbol{\tilde \nu_i}$}   
	\put(133,54){\small$\boldsymbol{\bar d_k}$} 
	\put(80,50){\small$\boldsymbol{(d)}$} 
    \end{picture}}
\vspace{-1cm}
\caption{Feynman diagrams inducing transitions of the type
  $\tau^-\to\mu^-\bar uu$ (a) or $\tau^-\to\mu^-\bar dd$ (b)--(d)
  by exchange of supersymmetric particles. The arrows denote the
  direction of the flow of the left-handed fields.
The particle labels denote the incoming ($\tau^-$) or outgoing fields
(all others). Thus in (b), e.g., we have an outgoing (SU(2) singlet) $d_j$ and 
(SU(2) singlet) $\bar d_\ell$. \label{fig:diags}}
\end{figure}
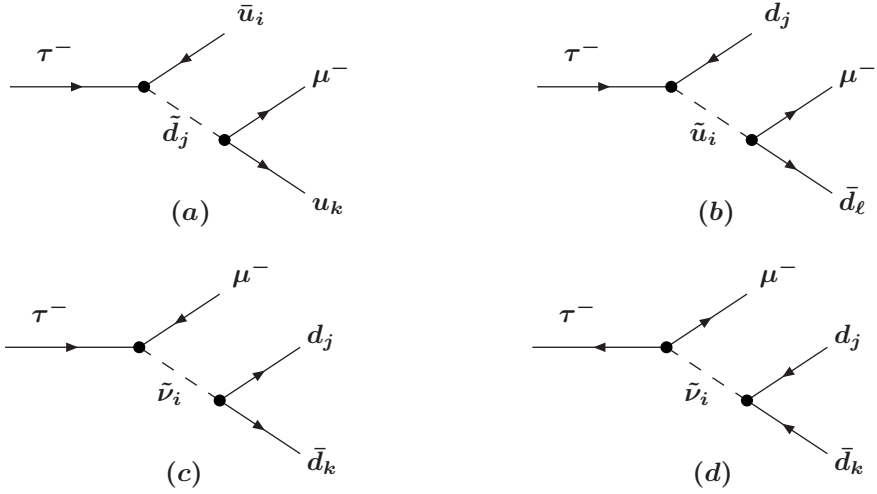
\item[(d)] Combining
$\epsilon_{ab} \lam_{2i3} L_2^a L_i^b  \bar E_3$
and $[\epsilon_{ab} \lam'_{ijk} L_i^aQ_j^b  \bar D_k]^\dagger$ corresponds
to $\tau^-\to\mu^-  \bar d_j d_k$ with 
\begin{equation}
\mathscr{L}_{\mathrm{eff}}=\frac{1}{2}
\frac{\lambda_{2i3}\lambda^{\prime *}_{ijk}}{m^2_{\tilde \nu_i}}
(\bar d_k P_L d_j)(\bar\mu  P_R \tau) ~.\label{eq:d}
\end{equation}
\end{enumerate}
Diagrammatic representations of contributions (a)--(d) are shown in Fig.~\ref{fig:diags}.

\section{Application to the vector current}

The invariant mass distribution of the width for the decay $\tau\to \mu \pi^+\pi^-$ is
given by
\begin{equation}
\frac{d\Gamma}{d\sqrt{s}} =\frac{1}{(2\pi)^5}\frac1{16m_\tau^2}\,{\overline{\left|
      \M\right|^2}}|\vec{p}_{\pi^+}^{\, *}|d\Omega_+^*\,|{\vec p}_\mu|d\Omega_\mu \ ,
\end{equation}
where $s$ is the invariant mass squared of the pion pair, 
$|{\vec p}_{\pi^+}^{\, *}|$ ($\Omega_{\pi^+}^*$) is the momentum (angle) of the ${\pi^+}$
in the rest frame of the pion pair, and $|{\vec p}_\mu|$ and $\Omega_\mu$
are to be given in the $\tau$ rest frame.

The essential observation is that the matrix element $\M$ factorizes,
since the primary transition is short-ranged (the range of interaction is
set by the inverse mass of the exchanged supersymmetric particles), while the
final-state interaction is  long-ranged. Therefore, we
may write
\begin{equation}
\label{factorization}
\M = \big\langle \pi^+\pi^-\big|\hat O^{(\alpha)}\big|0\big\rangle M^r_{(\alpha)} \ .
\end{equation}
The reduced matrix elements $M^r_{(\alpha)}$ are to be calculated in the underlying,
fundamental theory, while the hadronic matrix elements 
$\big\langle \pi^+\pi^-\big|\hat O^{(\alpha)}\big|0\big\rangle$
can be deduced either from data or determined with theoretical input.

To begin with, let us assume that only the vector current contributes
here---the generalization to the scalar current is straightforward and
will be presented below.  As long as isospin is assumed to be
conserved, two pions with vector quantum numbers (i.e., in a $P$-wave)
only couple to the {\it isovector} component of the current.
Therefore we need to consider only a single form factor and the
hadronic matrix element is given by the pion vector form
factor, $F_V(s)$, defined via
\begin{equation}
\big\langle \pi^+(p_{\pi^+})\pi^-(p_{\pi^-})\big|\tfrac{1}{2}(\bar
u\gamma^\alpha u-\bar d\gamma^\alpha d)\big|0\big\rangle 
\equiv F_V(s)(p_{\pi^+}-p_{\pi^-})^\alpha \ .
\label{FVdef}
\end{equation}
$F_V(s)$ is very well known both from direct measurements of
$e^+e^-\to\pi^+\pi^-$~\cite{Na7FF,KLOE-1,CMD2-1,CMD2-2,babarFF,KLOE-2}
and, via an isospin rotation, of $\tau^- \to \pi^-\pi^0\nu_\tau$~\cite{BelleFF}, 
as well as theoretical studies~\cite{Gasser:1990bv,guerrero,Oller,yndurain,Anant}.  It collects all
non-perturbative $\pi\pi$ interactions and is universal in the elastic
region, which to excellent approximation comprises the energy range
$s<1$\,GeV$^2$.\footnote{In the following section, we shall emphasize the importance of the 
large inelastic coupling of the pion--pion isospin $I=0$ $S$-wave to $K\bar K$ intermediate 
states in the region of the $f_0(980)$ resonance.  We wish to point out here that the coupling 
to kaons has, in contrast, an entirely negligible effect on the pion--pion $P$-wave.
In this case, the inelasticity is dominated by $4\pi$ intermediate states, often thought to 
be effectively clustered as $\pi\omega$ (compare Ref.~\cite{omega3pi}), and only rises very slowly
roughly above the $\pi\omega$ threshold.}

\begin{figure}
  \begin{center}
    \includegraphics[width=0.6\linewidth]{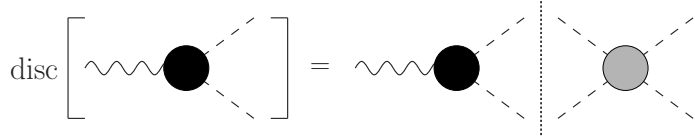} 
    \caption{Graphical representation of the discontinuity relation for the pion form factor.
    The dashed lines are pions, while the wavy line stands for the current coupling to these.
    The black disc denotes the form factor, while the gray disc denotes the pion--pion scattering $T$-matrix
    (which, by angular momentum conservation, is automatically projected onto the appropriate partial wave).}
    \label{fig:FFunit}
  \end{center}
\end{figure}
We briefly illustrate how to describe $F_V(s)$ theoretically, based solely on the fundamental principles of 
analyticity and unitarity.  Figure~\ref{fig:FFunit} gives a graphical illustration of the discontinuity
of the form factor, regarded as an analytic function of $s$ in the complex plane
cut along (parts of) the positive real axis, in the elastic regime, 
i.e., considering two-pion intermediate states only.  It is given by
\begin{align}
\text{disc}\, F_V(s) &= F_V(s+i\epsilon)-F_V(s-i\epsilon) = 2i\, \Im\, F_V(s) \nnnl
&
= 2i\, F_V(s) \,\sigma(s)\,  t_1^1(s)^* \, \theta \big(s-4M_\pi^2\big)
= 2i\, F_V(s) \sin\delta_1^1(s) e^{-i\delta_1^1(s)} \,\theta \big(s-4M_\pi^2\big) \,, \label{eq:disc}
\end{align}
where $\sigma(s) = \sqrt{1-4M_\pi^2/s}$ is proportional to the two-particle phase space, and
$t_\ell^I(s)$ refers to the pion--pion partial-wave amplitude of isospin $I$ and angular momentum $\ell$,
obtained from expanding the corresponding $T$-matrix in Legendre polynomials.
In the final step, we have rewritten the $P$-wave amplitude in terms of the phase shift $\delta_1^1(s)$ in the 
canonical manner.  One immediately deduces Watson's final-state theorem~\cite{watson}:
reality of $\Im\, F_V(s)$ implies that the phase of $F_V(s)$ coincides with $\delta_1^1(s)$.
The solution to Eq.~\eqref{eq:disc} is given by
\begin{equation}
F_V(s) = P_V(s) \Omega_1^1(s) ~, \quad
\Omega_1^1(s) =\exp\bigg\{\frac{s}{\pi}\int_{4M_\pi^2}^\infty ds'\frac{\delta_1^1(s')}{s'(s'-s)} \bigg\} ~,
\label{eq:Omnes}
\end{equation}
where $\Omega_1^1(s)$ is the Omn\`es function~\cite{Omnes} and $P_V(s)$ a polynomial.
The pion--pion phase shifts are known to excellent precision (up to at least $\sqrt{s} \simeq 1.1$\,GeV) from analyses
of the highly constrained system of dispersion relations known as Roy equations~\cite{Roy,ACGL,CCL,garcia}.

Perturbative QCD suggests $F_V(s)$ should fall off like $1/s$ for large values of $s$
(up to logarithmic corrections)~\cite{BrodskyLepage}, which is also the behavior of the Omn\`es function if the phase shift 
approaches $\pi$ asymptotically, as phenomenology indeed suggests.  Hence, $P_V(s)$ is required to be a constant.
Gauge invariance finally requires the normalization to be fixed to $F_V(s=0)=1$, 
therefore $P_V(s) \equiv 1$. 
The representation Eq.~\eqref{eq:Omnes} can be improved by taking inelastic effects (beyond two-pion intermediate states)
into account.  We here use a parametrization presented in Ref.~\cite{FFnew},
which describes the available high-precision data~\cite{Na7FF,KLOE-1,CMD2-1,CMD2-2,babarFF,KLOE-2,BelleFF} perfectly.

The relevant reduced matrix elements needed to complete Eq.~\eqref{factorization}---read off from the effective
Lagrangians given in Eqs.~\eqref{eq:a}--\eqref{eq:d}---can be subsumed
in the expression
\begin{equation}
M^r_{V\alpha} = \lambda_V \left[\bar \mu(p_\mu)\gamma_\alpha P_L \tau(p_\tau)\right]  \ , 
\label{mred_V}
\end{equation}
where the effective coupling $\lambda_V$ is given by
\begin{equation}
\lambda_V \equiv \frac{1}{4}\bigg(
\frac{\lambda'_{31j}\lambda^{\prime *}_{21j}}{m_{\tilde d_j}^2} - \frac{\lambda'_{3i1}\lambda^{\prime *}_{2i1}}{m_{\tilde u_i}^2}
\bigg) ~. \label{eq:matchLV}
\end{equation}
This in principle comprises six different contributions ($i,j=1,2,3$)
which can enhance each other or lead to cancellations, depending on
the phases of the R-parity-violating couplings. In the Standard Model
there is a strong hierarchy among the Yukawa couplings. For example
the top quark Yukawa coupling is almost a factor of forty larger than
the bottom quark Yukawa coupling.  Since no supersymmetry with or
without R-parity has yet been found, we shall for simplicity consider
one product of operators at a time in Eq.~\eqref{eq:matchLV}. We thus
employ the assumption that there is a hierarchy in the unknown
R-parity-violating
couplings~\cite{Barger:1989rk,Dreiner:1991pe}.\footnote{We
    keep in mind that Nature does not necessarily obey such
    analogies. For example the PMNS neutrino mixing angles are large
    compared to the CKM quark mixing angles.}

\begin{sloppypar}
Using Eqs.~\eqref{FVdef} and \eqref{mred_V} in the definition of the matrix element we find for
the spin-averaged squared matrix element
\begin{equation}
\frac{1}{2}\sum_{\rm spins} |\M|^2 = \frac{\lambda_V^2}{2} \big(s-4M_\pi^2\big)
\bigg[m_\tau^2+m_\mu^2-s
 + \frac{\lambda(m_\tau^2,m_\mu^2,s)}{s} z^2 \bigg] 
\big|F_V(s)\big|^2 ~,
\label{Msquared}
\end{equation}
where we included a prefactor of 1/2 from averaging the incoming
$\tau$ polarizations. $\lambda(x,y,z)=x^2+y^2+
z^2-2(xy+xz+yz)$  is the usual K\"all\'en function,
and $z=\cos\theta$ denotes the angle of the pions relative to the 
leptons in the $\pi\pi$ rest frame. Evaluating the angular integral
and collecting all kinematic prefactors,  we arrive at
\begin{equation}
\frac{d\Gamma_V}{d\sqrt{s}} = \frac{\lambda_V^2}{256\pi^3m_\tau^3}
\big(s-4M_\pi^2\big)^{3/2}
\lambda^{1/2}\big(m_\tau^2,m_\mu^2,s\big) 
\bigg[m_\tau^2+m_\mu^2-s + \frac{\lambda(m_\tau^2,m_\mu^2,s)}{3s}  \bigg] 
\big|F_V(s)\big|^2\ .
\label{Vdist_final}
\end{equation}
Since all quantities but $\lambda_V$
are known in Eq.~\eqref{Vdist_final}, it provides the kind of expression we are
looking for. In particular, the norm of the hadronic current is fixed
unambiguously in contrast to previous studies, where quark-model wave functions
were employed to fix the normalization. 
\begin{figure}
\begin{center}
\includegraphics*[width=\linewidth]{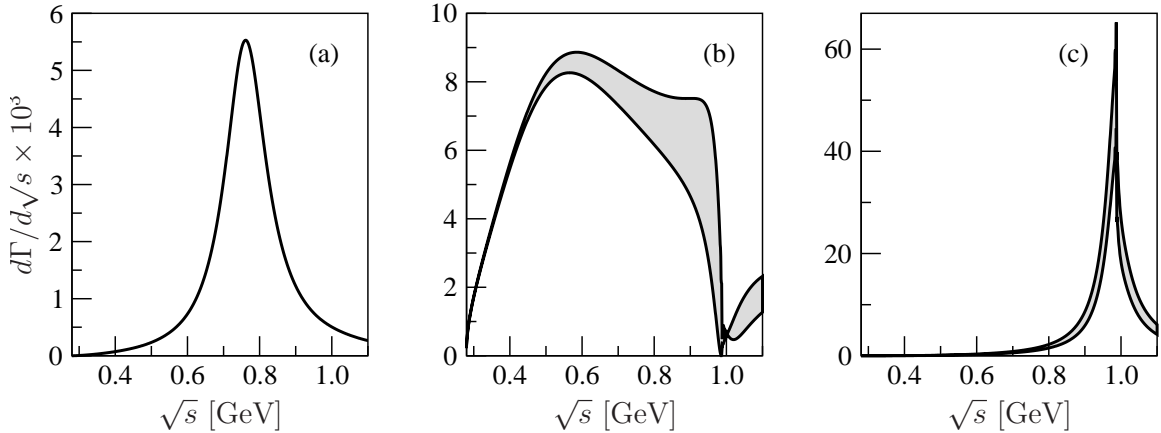}
\caption{The predicted signals individually for the currents  (a) 
$\bar q\gamma^\alpha q$, (b) $(\bar uu+\bar dd)/2$, and (c)~$\bar s s$. In all 
cases the effective coupling constant is set to 1\,GeV$^{-2}$. 
For the uncertainty bands of the scalar form factors, see the discussion in the main text.
\label{fig:differentialrates}}
\end{center}
\end{figure}
The resulting $\sqrt{s}$ distribution is depicted in panel (a) 
of Fig.~\ref{fig:differentialrates}, where we have used
$\lambda_V = 1$\,GeV$^{-2}$, which is easily rescaled.
\end{sloppypar}

The signal in the vector channel can be represented reasonably well by a
Breit--Wigner distribution with an energy-dependent width as provided by various
parametrizations. Here the advantage of our approach lies mainly in its
fixed normalization---we briefly compare to the narrow-resonance approximation (where one studies
the decay $\tau\to\mu\,\rho^0(770)$, assuming a stable $\rho^0$) in Appendix~\ref{app:narrow}.
An additional advantage is
that not only the modulus of the form factor, but also its phase is fixed
unambiguously, such that the interference of various currents can be analyzed
as well.

\section{Application to the scalar currents}

Contrary to the vector currents, scalar currents are typically not
well represented by Breit--Wigner functions (for a detailed discussion
see e.g.\ Ref.~\cite{Gardner:2001gc}).  In this case, isospin symmetry
requires two pions to couple exclusively to {\it isoscalar} scalar
currents.  However, there are two such form factors, originating from
non-strange and strange isoscalar scalar sources. They are called
$\Gamma_\pi^n$ and $\Gamma_\pi^s$, respectively, and will contribute
simultaneously.  The use of scalar form factors (as opposed to
Breit--Wigner parametrizations) is unavoidable if one wants to
determine the underlying couplings in a controlled way.

The expressions analogous to Eq.~\eqref{FVdef} now read
\begin{equation} \left\langle \pi^+\pi^-\left|\bar qq\right|0\right\rangle \equiv
\B^q\Gamma_\pi^q(s) \ ,
\label{Fsdef}
\end{equation}
where the quark flavors may be either $\bar qq=(\bar uu+\bar dd)/2$
for the light quarks, with the superscript $q=n$ denoting the
corresponding scalar form factor, or $\bar qq=\bar ss$ for strange
quarks (with superscript $q=s$).  Furthermore,
$\B^n=M_\pi^2/(m_u+m_d)$, $\B^s=(2M_K^2-M_\pi^2)/(2m_s)$.  With this
convention, the form factors, $\Gamma^q_\pi(s)$, are invariant under the QCD
renormali\-zation group.  For the numerical evaluation, we will use the
values obtained from averaging lattice computations with $N_f=2+1$
dynamical flavors~\cite{FLAG}, $(m_u+m_d)/2=(3.43\pm0.11)\,$MeV,
$m_s=(94\pm3)\,$MeV, to be understood in the $\overline{\rm MS}$
scheme at the running QCD scale $\mu=2\,$GeV.  In addition, in analogy
to Eq.~\eqref{mred_V} we may now write
\begin{equation}
M_S^{q \, r} = \lambda_S^q \left[\bar \mu(p_\mu) P_R \tau(p_\tau)\right]  \ ,
\label{mred_s}
\end{equation}
where, again, the superscript $q$ denotes the quark flavor fed by the
corresponding operator. Comparison to Eqs.~\eqref{eq:c} and \eqref{eq:d} yields
\begin{equation}
\lambda_S^n \equiv \frac{(\lambda_{3i2}+\lambda_{2i3})\lambda^{\prime *}_{i11}}{4m^2_{\tilde \nu_i}} ~, \quad
\lambda_S^s \equiv \frac{(\lambda_{3i2}+\lambda_{2i3})\lambda^{\prime *}_{i22}}{4m^2_{\tilde \nu_i}} ~.
\end{equation}
With these expressions we find
\begin{equation}
\frac{d\Gamma_S}{d\sqrt{s}} = \frac{1}{64\pi^3m_\tau^3}
\big(s-4M_\pi^2\big)^{1/2}
\lambda^{1/2}\big(m_\tau^2,m_\mu^2,s\big)
\Big[ (m_\tau+m_\mu)^2-s \Big] 
\big|\B^n\lambda_S^n\Gamma_\pi^n(s)+\B^s\lambda_S^s \Gamma_\pi^s(s)\big|^2 \,.
\label{Sdist_final}
\end{equation}

Experimentally, the scalar form factors are not accessible as directly
and unambiguously as the vector form factor.  However, they can be
reconstructed from dispersion theory, similar to what we discussed as
the Omn\`es representation of the vector form factor in the previous
section.  The main difference is that the elastic approximation breaks
down much earlier in the pion--pion $S$-wave (of isospin $I=0$) due to
the strong inelastic coupling of two $S$-wave pions to $K\bar K$ in
the region of the $f_0(980)$, i.e., beginning immediately at the $K\bar K$ threshold.
In order to describe the scalar form factors including energies around
the mass of the $f_0(980)$, it is therefore mandatory to solve a
two-channel Muskhelishvili--Omn\`es
problem~\cite{Omnes,Muskhelishvili}.  The discontinuity
equation~\eqref{eq:disc} has to be generalized to two coupled channels
for the scalar form factors of pion and kaon.  The two-channel
$T$-matrix correspondingly can be parametrized in terms of
\textit{three} input functions, the pion--pion $S$-wave phase shift
$\delta_0^0(s)$ known from Roy equation
solutions~\cite{ACGL,CCL,garcia}, as well as modulus and phase of the
inelastic reaction $\pi\pi\to K\bar K$~\cite{Cohen,Etkin,BDM04}.  We
again assume the fall-off of all scalar form factors $\propto 1/s$,
and the scattering phases involved to approach integer multiples of
$\pi$ in the appropriate way.  The solution to the coupled-channel
discontinuity equation cannot be written down analytically in a similarly 
compact form as for the single-channel case, Eq.~\eqref{eq:Omnes},
but has to be constructed
numerically~\cite{DGL90,Moussallam99,Descotes,HDKM}.  It now depends
on the constant normalization of the corresponding pion {\it and} kaon
form factors $\Gamma_{\pi/K}^{n/s}(s=0)$.  In contrast to the vector
case, the normalizations of the scalar form factors are not fixed by
symmetries; they are however related to the corresponding masses by
the Feynman--Hellmann theorem~\cite{Hellmann,Feynman},
\begin{equation}
\Gamma_\pi^n(0) = \frac{1}{2\B^n}\bigg(\frac{\partial}{\partial m_u}+\frac{\partial}{\partial m_d}\bigg) M_\pi^2 ~,\quad
\Gamma_\pi^s(0) = \frac{1}{\B^s}\frac{\partial}{\partial m_s}M_\pi^2 ~,
\end{equation}
and similar for the scalar form factors of the kaon.
At leading order in the quark mass expansion, one therefore has $\Gamma_\pi^n(0)=1$, $\Gamma_\pi^s(0)=0$, 
$\Gamma_K^n(0)=1/2$, and $\Gamma_K^s(0)=1$. 
Beyond that, information on these quantities can again be deduced from lattice calculations.
We vary the normalizations of the kaon form factors according to
$\Gamma_K^n(0)=0.4\ldots0.6$, $\Gamma_K^s(0)=0.95\ldots1.15$, as suggested by the uncertainties in 
the corresponding low-energy constants given in Ref.~\cite{FLAG}, 
while keeping the rather well-known pion-form-factor normalizations fixed at
$\Gamma_\pi^n(0)=0.98$, $\Gamma_\pi^s(0)=0$.\footnote{A recent analysis of these scalar form factor
normalizations directly based on lattice data, following the generalized framework of \textit{resummed chiral perturbation
theory} as in Ref.~\cite{Toucas}, yields  
$\Gamma_\pi^n(0)=1.000\pm0.005$, 
$\Gamma_\pi^s(0)=0.013\pm0.009$, 
$\Gamma_K^n(0)=0.56\pm0.06$, and 
$\Gamma_K^s(0)=1.19\pm0.11$, 
thus perfectly compatible with the values assumed above, even though still more precise in some cases.
We are very grateful to V\'eronique~Bernard and S\'ebastien~Descotes-Genon for communicating these results to us
prior to publication.}

In Figs.~\ref{fig:differentialrates}(b) and (c) we show the resulting
invariant mass distributions for the pion pair for
$(\lambda_S^q=1$\,GeV$^{-2}$, $\lambda_S^s=0$) and $(\lambda_S^q=0$,
$\lambda_S^s=1$\,GeV$^{-2}$), respectively, with the uncertainty bands
as dictated by the above estimates for the uncertainty in the kaon
form factor normalizations.  In panel $(b)$ the $f_0(500)$ (or
$\sigma$ meson) shows up as a broad bump with a clear
non-Breit--Wigner shape, while the $f_0(980)$ produces a peak
exclusively in the strangeness form factor, panel $(c)$.  Thus, were a
pronounced peak just below 1\,GeV observed in $\tau\to \mu \pi^+\pi^-$,
it would allow one to straightforwardly extract $\lambda_S^s$ from
data, without the need to employ any assumption on the internal
structure of the $f_0(980)$---additional information can be gained
from also studying the $\bar KK$ final state, which however we will
not detail here. This highlights the advantage of our approach compared to
the one of  Ref.~\cite{herrero09}: in that work assumptions
on the quark content of the $f_0(980)$ need to be employed in order to derive
bounds, which then in turn strongly depend on these assumptions. In our
case the bounds can be deduced directly from a fit to the spectra, once
they are measured.

\section{Discussion}

If supersymmetry was to show up in experiments like $\tau \to \mu \pi^+\pi^-$, 
there is no a priori reason why pion pairs in the vector channel would be
significantly more populated than pion pairs in the scalar channel.
While the effective couplings for the vector channel are given by squark exchange,
the scalar channel is driven by sneutrino-exchange contributions.
Thus one should expect interferences of the three
currents discussed individually above. In this context it is important to
stress that, below the first significant inelastic threshold, the phase of 
the form factor agrees with that of the elastic scattering amplitude~\cite{watson}, which
is well known in both the scalar and the vector channel~\cite{ACGL,CCL,garcia}.

The Belle collaboration has given upper limits on branching ratios 
$\B(\tau^-\to\mu^- \pi^+\pi^-)$ with different kinematical 
cuts~\cite{Miyazaki:2008mw,Miyazaki:2011xe,Miyazaki:2012mx}.
In particular, they find
\begin{align}
\B\big(\tau^-\to\mu^-\rho^0(770)\big)\times\B\big(\rho^0(770)\to \pi^+\pi^-\big) &< 1.2\times 10^{-8} ~,\nnnl
\B\big(\tau^-\to\mu^-f_0(980)\big)\times\B\big(f_0(980)\to \pi^+\pi^-\big) &< 3.4\times 10^{-8} ~,\nnnl
\B\big(\tau^-\to\mu^- \pi^+\pi^-\big) &< 2.1\times 10^{-8} ~. \label{eq:expbounds}
\end{align}
The resonance signals are isolated by applying cuts to the $\pi^+\pi^-$ 
invariant mass spectrum, specifically $906\,{\rm MeV} <\sqrt{s} < 1065\,{\rm MeV}$ for the 
$f_0(980)$~\cite{Miyazaki:2008mw} and $587\,{\rm MeV} < \sqrt{s}<962\,{\rm MeV}$ 
for the $\rho(770)$~\cite{Miyazaki:2011xe}.  As we aim
for deriving upper limits on coupling constants from null experiments,
what will effectively enter the bounds on the scalar couplings are the
\textit{lower} limits of the uncertainty bands on the scalar form
factors.  Note that the vector form factor is known to a precision that any
uncertainty therein is totally irrelevant at the accuracy we aim for.
When comparing to the last, inclusive, bound in
Eq.~\eqref{eq:expbounds}, we will set all form factors to zero above
$1.1$\,GeV where we deem our representations not very reliable
anymore; obviously, the bounds could be improved upon if lower bounds
were available also at higher energies.  Integrating
Eqs.~\eqref{Vdist_final} and \eqref{Sdist_final} in the respective
ranges, we find
\begin{align}
\int_{587\,{\rm MeV}}^{962\,{\rm MeV}}\bigg(\frac{d\Gamma_V}{d\sqrt{s}}&+\frac{d\Gamma_S}{d\sqrt{s}}\bigg)d\sqrt{s} 
= \Big[0.94|\lambda_V|^2 + \{2.3\ldots3.0\}|\lambda_S^n|^2 + \{2.5\ldots3.7\}\,\Re\big(\lambda_S^n\lambda_S^{s*}\big) \nnnl 
&   + \{1.2\ldots1.8\}|\lambda_S^s|^2\Big] \times 10^{-3} {\rm GeV}^5  \nnnl
& < 2.7\times 10^{-20} {\rm GeV} \ , \nnnl
\int_{906\,{\rm MeV}}^{1065\,{\rm MeV}}\bigg(\frac{d\Gamma_V}{d\sqrt{s}}&+\frac{d\Gamma_S}{d\sqrt{s}}\bigg)d\sqrt{s} 
= \Big[0.10|\lambda_V|^2 + \{0.3\ldots0.6\}|\lambda_S^n|^2 + \{0.5\ldots2.1\}\,\Re\big(\lambda_S^n\lambda_S^{s*}\big)\nnnl 
& + \{0.35\ldots0.42\}\,\Im\big(\lambda_S^n\lambda_S^{s*}\big) + \{2.5\ldots3.7\}|\lambda_S^s|^2\Big] \times 10^{-3} {\rm GeV}^5 \nnnl
& < 7.7\times 10^{-20} {\rm GeV} \ , \nnnl 
\int_{2M_\pi}^{1100\,{\rm MeV}}\bigg(\frac{d\Gamma_V}{d\sqrt{s}}&+\frac{d\Gamma_S}{d\sqrt{s}}\bigg)d\sqrt{s} 
= \Big[1.05|\lambda_V|^2 + \{4.3\ldots5.1\}|\lambda_S^n|^2 + \{2.4\ldots4.7\}\,\Re\big(\lambda_S^n\lambda_S^{s*}\big)\nnnl
& + \{0.47\ldots0.57\}\,\Im\big(\lambda_S^n\lambda_S^{s*}\big) + \{3.2\ldots4.7\}|\lambda_S^s|^2\Big] \times 10^{-3} {\rm GeV}^5 \nnnl
& < 4.8\times 10^{-20} {\rm GeV} \ ,
\end{align}
where the limits on the branching fractions have been converted into
partial widths with the help of the lifetime of the $\tau$ lepton.
The uncertainties shown for the coefficients of the scalar couplings
are due to the ranges assumed for the kaon scalar form factor
normalizations as discussed above. They are displayed explicitly in
order to indicate the remaining potential for improvement, once this
specific hadronic input is still better known.  Note that
contributions $\propto \Im\big(\lambda_S^n\lambda_S^{s*}\big)$ have to
come from above the $K\bar K$ threshold, where the two interfering
scalar form factors are not required to have identical phases
according to Watson's theorem~\cite{watson} any more.

In order to set limits on the underlying coupling constants, we shall make the usual simplifying assumption
that they are all real.
Thus, assuming the first operator in Eq.~(\ref{eq:matchLV}) dominates we obtain for example the bound
\begin{equation}
\lam_{21i}^{\prime*}\lam_{31i}^\prime<2.1\cdot10^{-4}\left(\frac{m_{\tilde d_i}}{100\,\mathrm{GeV}}\right)^2\,.
\end{equation} 
The corresponding bound on the second operator is
\begin{equation}
\lam_{2i1}^{\prime*}\lam_{3i1}^\prime <2.1\cdot10^{-4}\left(\frac{m_{\tilde u_i}}{100\,\mathrm{GeV}}\right)^2\,.
\end{equation} 
A complete list is given in Table~\ref{bounds}. 
\begin{table}
\begin{center}
\renewcommand{\arraystretch}{1.2}
\begin{tabular}{cccc}
\toprule
product of couplings  & bound & susy mass & eff.\ coupling \\
\midrule
$\lam_{21i}^{\prime*}\lam_{31i}^\prime$ & $2.1\cdot10^{-4}$ &$m_{\tilde d_i}$ & $\lam_V$ \\ 
$\lam_{2i1}^{\prime*}\lam_{3i1}^\prime$ & $2.1\cdot10^{-4}$ &$m_{\tilde u_i}$ & $\lam_V$ \\ 
$\lam_{3i2}\lam_{i11}^{\prime*}$,
$\lam_{2i3}\lam_{i11}^{\prime*}$ & $1.3\cdot10^{-4}$ & $m_{\tilde \nu_i}$ & $\lam_S^n$ \\
$\lam_{3i2}\lam_{i22}^{\prime*}$,
$\lam_{2i3}\lam_{i22}^{\prime*}$ & $1.5\cdot10^{-4}$ & $m_{\tilde \nu_i}$ & $\lam_S^s$ \\
\bottomrule
\end{tabular}
\renewcommand{\arraystretch}{1.0}
\end{center}
\caption{Our best bounds on the products of supersymmetric R-parity-violating couplings. 
  We assume all other couplings to be zero. The
  bounds scale with the square of the given mass divided by  100\,GeV, so for a 1\,TeV 
  mass they are a factor 100 weaker. To obtain the bound we 
employed the effective coupling in the far right column. Note that in the first
two rows for the case of $i=1$ both operator products contribute in
$\lam_V$ and the relative phases are essential. They can only be
treated separately if $m_{\tilde d_1}\ll m_{\tilde u_1}$ or $m_{\tilde  u_1}\ll m_{\tilde d_1}$.
\label{bounds}}
\end{table}
Note that while the most restrictive bound on the vector coupling
$\lambda_V$ and the underlying products of fundamental coupling
constants is derived from the dedicated search for
$\tau^-\to\mu^-\rho^0(770)$~\cite{Miyazaki:2011xe}, the best limits on
the two effective scalar couplings stem from the most recent limit on
$\tau^-\to\mu^-\pi^+\pi^-$ without further kinematical
cuts~\cite{Miyazaki:2012mx}.  In the literature the best
previous bound is given from different processes bounding the
individual couplings separately~\cite{Allanach:1999ic}. Combining them
to a product bound we obtain
\begin{equation}
\lam'_{21i}\lam'_{31i} < 7.2\cdot10^{-3} \left(\frac{m_{\mathrm{susy}}} {100\,\mathrm{GeV}}\right)^2\,,
\end{equation}
where we have set the mass of the different virtual supersymmetric scalars equal. We have thus improved this bound
by more than a factor of 30.  We can also compare our result to the related bounds obtained in Ref.~\cite{Black:2002wh}.
The authors consider the effective operators
\begin{equation}
\mathcal{O}_j=\frac{C^j_{\alpha\beta}}{\Lambda^2}(\bar\mu\Gamma_j\tau)(\bar
q^\alpha\Gamma_j q^\beta)\,,
\end{equation}
independent of their origin, for different Dirac structures $\Gamma_j$.
These contribute for example to
$\tau\to\mu\pi^+\pi^-$. Setting $C^j=4\pi$ the authors obtain the bounds 
\begin{equation}
\Lambda > 2.6 \,\mathrm{TeV\; (scalar)}, \quad \Lambda>12\,\mathrm{TeV\; (vector)}\,,
\end{equation}
where `scalar' and `vector' denote the cases $\Gamma_j=\mathbf{1}$ 
and $\Gamma_j=\gamma^\alpha$, respectively. Comparing the vector case to our bound we thus have
\begin{equation}
\frac{\lam'_{21i}\lam'_{31i}}{4\tilde m^2} = \frac{4\pi}{\Lambda^2} \,,
\end{equation}
hence
\begin{equation}
\lam'_{21i}\lam'_{31i} = 16\pi\left(\frac{\tilde m}{\Lambda}\right)^2
< 3.5\cdot10^{-3} \left(\frac{\tilde m}{100\,\mathrm{GeV}}\right)^2\,,
\end{equation}
which is more than an order of magnitude weaker than our bound, partly
owing to weaker experimental bounds at the time.  We point out that in
Ref.~\cite{Black:2002wh} the scalar form factors were assumed to be
constant, which is, as we have discussed, not a good approximation.

\section{Conclusions}

Supersymmetry has to-date not been observed. Thus well-motivated
versions, such as R-parity violation, which however have been
considered less conventional, are now also being investigated in more
detail. Most of the strictest bounds on supersymmetry with R-parity
violation arise from precision, low-energy observables, e.g.\ meson
decays, or decays involving mesons. These computations typically
involve simple approximations of the relevant hadron physics, such as
the current algebra approximation. In this paper we demonstrate that
the bounds on the R-parity-violating couplings can be considerably
improved when including dynamical aspects of  hadron physics.
 We do this in a model-independent manner, such that it can
  easily be applied to test other fundamental theories of physics beyond
  the Standard Model.

  To be specific, we have focused on the decay
  $\tau^\pm\to\mu^\pm\pi^+\pi^-$.  Here, the hadron physics aspects
  can be treated particularly rigorously, as the strong final-state
  interactions of the pion pair are described in terms of the vector
  and scalar form factors, which are either directly measured
  experimentally, or can be reconstructed using the rigorous,
  model-independent methods of dispersion theory. We then employ upper
  bounds found by the Belle collaboration on the branching ratios of
  the lepton-flavor-violating $\tau$ decays, and obtain bounds on
  products of the R-parity-violating couplings. Due to the extra information we have
  included and the slightly improved experimental data we find bounds
  which are more than an order of magnitude stronger than previous
  ones.

\section*{Acknowledgments}
We are grateful to Gilberto Colangelo for providing us with yet unpublished results of Ref.~\cite{CCL},
and Bachir Moussallam for providing a version of the $\pi\pi\to K\bar K$ amplitudes
consistent with Ref.~\cite{CCL}.
Furthermore, we would like to thank Martin Hoferichter for useful discussions.
One of us (HKD) would like to thank the Aspen Center for Physics,
where part of this work was completed.

\appendix

\section{Derivation of effective operators}\label{app:effops}

In this Appendix we present as an example the computation of the
effective Lagrangian given in Eq.~\eqref{eq:a}. We show the results in
four-component spinor notation, as this still is the most widely used
convention across communities. However, we actually derived the
results in two-component spinor notation~\cite{Dreiner:2008tw} and
then translated them into four-component notation using Appendix~G of
Ref.~\cite{Dreiner:2008tw}.

The superpotential in Eq.~\eqref{rpv-superpot} contains the terms $
\epsilon_{ab}\lambda_{ijk}^\prime L_i^a Q_j^b \bar D_k$. In component fields
the corresponding Yukawa Lagrangian terms are given by, see for
example Ref.~\cite{Richardson:2000nt},
\begin{align}
\mathscr{L}_{\mathrm{LQD}} =-\lambda_{ijk}^\prime\Big(
&\tilde d^*_{kR} \overline{\nu^c_i}P_Ld_j 
-\tilde d^*_{kR} \overline{\ell^c_i}P_Lu_j 
+\tilde d_{jL}\bar d_k P_L \nu_i \nnnl
& -\tilde u_{jL} \bar d_k P_L \ell_i
+\tilde\nu_i\bar d_kP_L d_j
-\tilde\ell_{iL}\bar d_k P_L u_j
+\mathrm{h.c.}\Big)\,,\label{explic-lagr}
\end{align}
where we have assumed that the Yukawa coupling is real and we have
denoted the charged lepton by $\ell_i$. The charge conjugate field is
defined as $\ell_i^c\equiv C{{\bar\ell_i}}^T$. If we now combine the
second term for $i=3$ as well as the hermitian conjugate of the second 
term for $i=2$, identify the third index, and assume the
intermediate $\tilde d_k$ squark to be very heavy, we obtain the effective 
Lagrangian
\begin{equation}
\mathscr{L_\mathrm{eff}}=-\frac{\lambda'_{3ij}\lambda_{2kj}^{\prime*}}{m^2_{\tilde
  d_j}}   \left(\overline{\tau^c} P_L u_i\right) \left( \bar u_k P_R \mu^c
\right) ~.
\end{equation}
A Fierz reordering as well as employing the definition of the complex
conjugate field then gives the form for the effective interaction in Eq.~\eqref{eq:a}
\begin{equation}
\mathscr{L}_{\mathrm{eff}}=\frac{1}{2}
\frac{\lambda'_{3ij}\lambda^{\prime *}_{2kj}}{m^2_{\tilde d_j}}
(\bar u^k\gamma^\alpha P_L u^i)(\bar\mu \gamma_\alpha P_L \tau) ~.
\end{equation}

\section{Comparison to the narrow-resonance approximation}\label{app:narrow}

In this Appendix, we briefly compare our treatment of the $\tau \to \mu \pi^+\pi^-$ decay 
in terms of form factors to the narrow-resonance approximation.
If, e.g., the $\rho$ meson were a stable particle, instead of
Eq.~\eqref{factorization} we had to use, cf.\ Ref.~\cite{herbi02,herbi07},
\begin{equation}
\label{factorization_stable}
\M = \big\langle \rho (k)\big|\hat J^{\alpha}\big|0\big\rangle
M^r_{V \, \alpha} \ , \quad
\big\langle \rho (k)\big|\hat J^{\alpha}\big|0\big\rangle=\big\langle \rho(k)\big|\tfrac{1}{2}(\bar u\gamma^\alpha u-\bar d\gamma^\alpha d)\big|0\big\rangle =
\frac{1}{\sqrt{2}} f_\rho M_\rho \epsilon(\lambda)^{\alpha \ *} \ ,
\end{equation}
where $f_\rho=220$\,MeV denotes the $\rho$ decay constant and $\epsilon(\lambda)$ is
the $\rho$ polarization vector. The expressions
for the physical and the stable $\rho$ are most easily compared on the level
of the decay rates. Employing the identity
\begin{equation}
\sum_\lambda  \epsilon(\lambda)_\alpha^*  \epsilon(\lambda)_\beta
=-g_{\alpha \beta} +  \frac{k_\alpha k_\beta}{M_\rho^2} \ ,
\end{equation}
one finds that switching from a stable $\rho$ to an unstable one
is obtained by the replacement
\begin{align}
\frac{1}{2}f_\rho^2 M_\rho^2 &\to -\frac13 (2\pi)^3\int ds \int
d\Phi_2(q,p_{\pi^+},p_{\pi^-})({p}_{\pi^+}-{p}_{\pi^-})^2\left|F_V(s)\right|^2 \nnnl
&= \frac{1}{6\pi^2}\int ds \Theta(s-4M_\pi^2)\frac{|{\vec p}_{\pi^+}^*|^3}{\sqrt{s}}\left|F_V(s)\right|^2 \ ,
\label{matching}
\end{align}
where $q^2=s$, we use the definition of the phase space as given
in Ref.~\cite{PDG}, and $({p}_{\pi^+}-{p}_{\pi^-})^2=-4{\vec p}_{\pi^+}^{\, * \ 2}$ in the center-of-mass frame.
In order to understand under which circumstances the full expression is
well represented by the left hand side of Eq.~\eqref{matching}, we may
parametrize the pion vector form factor by its spectral function, assuming
a constant width. Then we may write
\begin{equation}
|F_V(s)|^2 \simeq \frac{M_\rho^4}{(s-M_\rho^2)^2+M_\rho^2\Gamma_\rho^2} =
    \pi \frac{M_\rho^3}{\Gamma_\rho} \rho(s) \ .
\end{equation}
As a consequence of unitarity the spectral function is normalized,
\begin{equation}
\int_{4M_\pi^2}^\infty ds \,\rho(s) = 1 \ .
\end{equation}
In experiments the $\rho$ meson is typically identified by imposing cuts
on the two-pion invariant mass. If for the sake of the argument here we assume
that these cuts were sufficiently wide that the normalization integral
is exhausted, we find, using the explicit expression for the $\rho$
width
\begin{equation}
\Gamma_\rho = \frac{g_{\rho\pi\pi}^2}{6\pi}\frac{|{\vec p}_{\pi^+}^*|^3}{M_\rho^2} 
\end{equation}
as well as the KSFR relation $g_{\rho\pi\pi}^2=M_\rho^2/(2f_\pi^2)$~\cite{KS,FR},
\begin{equation}
f_\rho = 2 f_\pi \ ,
\end{equation}
which is a known connection e.g.\ in the hidden-local-symmetry approach~\cite{HLS}.
With $f_\pi=92.2$\,MeV, this identity is fulfilled to about 20\% accuracy. 
However, in any realistic situation the $\rho$ meson in the
final state can only be isolated by cuts on the $\pi\pi$ invariant
mass, such that the $\rho$ spectral function is not fully saturated,
leading to uncontrolled inaccuracies in the extraction of the effective parameters.


\end{document}